\documentclass[runningheads]{llncs}
\usepackage[T1]{fontenc}
\usepackage{graphicx}
\usepackage{booktabs}
\usepackage{multirow}
\usepackage[table,x11names]{xcolor}
\usepackage{hyperref}

\begin{document}
\title{Systematic Evaluation of Synthetic Data Augmentation for Multi-class NetFlow Traffic \thanks{Preprint; Accepted at Machine Learning for CyberSecurity @ European Conference on Machine Learning and Principles and Practice of Knowledge Discovery in Databases (ECML PKDD) 2024}}
\titlerunning{Synthetic Augmentation for NetFlow}
% If the paper title is too long for the running head, you can set
% an abbreviated paper title here
%
\author{
Maximilian Wolf\inst{1}
Dieter Landes\inst{1}
Andreas Hotho\inst{2}
Daniel Schlör\inst{2}
}
%\author{
%Anonymous Author
%}
%\author{First Author}
%
\authorrunning{M. Wolf et al.}
%\authorrunning{Author}

% First names are abbreviated in the running head.
% If there are more than two authors, 'et al.' is used.
%

\institute{Center for Responsible Artificial Intelligence, University of Applied Sciences and Arts Coburg, Friedrich-Streib-Str. 2 Coburg, Germany 
\email{\{maximilian.wolf, dieter.landes\}@hs-coburg.de} \and
Center for Artificial Intelligence and Data Science, University of Würzburg, Campus Hubland Nord, Emil-Fischer-Straße 50 Würzburg, Germany
\email{\{hotho, schloer\}@informatik.uni-wuerzburg.de}
}
\maketitle              % typeset the header of the contribution
\begin{abstract}
%The abstract should briefly summarize the contents of the paper in
%150--250 words.

The detection of cyber-attacks in computer networks is a crucial and ongoing research challenge. Machine learning-based attack classification offers a promising solution, as these models can be continuously updated with new data, enhancing the effectiveness of network intrusion detection systems (NIDS). Unlike binary classification models that simply indicate the presence of an attack, multi-class models can identify specific types of attacks, allowing for more targeted and effective incident responses. However, a significant drawback of these classification models is their sensitivity to imbalanced training data.

Recent advances suggest that generative models can assist in data augmentation, claiming to offer superior solutions for imbalanced datasets. Classical balancing methods, although less novel, also provide potential remedies for this issue. Despite these claims, a comprehensive comparison of these methods within the NIDS domain is lacking. Most existing studies focus narrowly on individual methods, making it difficult to compare results due to varying experimental setups.
To close this gap,  we designed a systematic framework to compare classical and generative resampling methods for class balancing across multiple popular classification models in the NIDS domain, evaluated on several NIDS benchmark datasets. 

Our experiments indicate that resampling methods for balancing training data do not reliably improve classification performance. Although some instances show performance improvements, the majority of results indicate decreased performance, with no consistent trend in favor of a specific resampling technique enhancing a particular classifier.

\end{abstract}

\section{Introduction}
Cybercrime and its detection and prevention is an ongoing topic of research.
A primary focus of attacks is the computer infrastructure of companies, which can be targeted in various ways.
The ENISA report of 2023 \cite{ENISA_2023} identifies key threats such as data theft, ransomware, and threats to system availability, including denial-of-service attacks and malware.
To detect network attacks, network intrusion detection systems (NIDS) are used to analyze a company's network traffic and identify malicious activities.
Previous studies have demonstrated the application of machine learning methods for the analysis and extraction of network traffic patterns, as highlighted in the comprehensive review by Salman et al.~\cite{Salman_Elhajj_Kayssi_Chehab_2020}.

In this work, we focus on network intrusion detection as a multi-class classification problem, where the classifier must distinguish between different types of attacks. 
In contrast to binary classification approaches, which only indicate whether an attack has occurred, a multi-class approach provides specific information about the type of attack, allowing for a tailored response to the specific threat, such as port scans, brute-force attacks or DDoS.
Since attack events are specific, but rare, their recording for benchmark purposes results in unbalanced class distributions. %\cite{} %datasets?
However, many standard multi-class models, such as Decision Trees, are sensitive to the class distribution and deliver low classification performance when trained on imbalanced datasets in other data domains~\cite{Haixiang_Yijing_Shang_Mingyun_Yuanyue_Bing_2017}.
Consequently,  multiple ways of handling imbalanced class distributions have been proposed, e.g. classification algorithms specifically for imbalanced learning~\cite{haixiang2017learning}, weighting schemes~\cite{steininger2021density}, cost-sensitive learning~\cite{elkan2001foundations} or resampling~\cite{Fernández_López_Galar_Del_Jesus_Herrera_2013}, which is particularly versatile as it is model-agnostic.

Resampling techniques, specifically under- and oversampling, balance the class distribution by drawing samples from the majority classes to match the number of entries in the minority class or vice versa. 
To further augment the training data, synthetic resampling methods such as SMOTE \cite{chawla2002smote} balance the dataset by generating synthetic data. % e.g. via linear interpolations.
These common approaches might introduce issues in high-dimensional and complex datasets, such as boundary bias~\cite{fernandez2018smote} and issues related to data-quality~\cite{gholampour2024impact,boudegzdame2024approach} as a linear feature  interpolation may not generate valid data points that accurately represent real-world scenarios, potentially leading to reduced classifier performance.
In contrast, conditional generative models such as Variational Autoencoders~\cite{kingma2013auto}, Generative Adversarial Networks~\cite{goodfellow2014generative} and diffusion-based models such as Latent-Diffusion~\cite{rombach2022high} promise to generate realistic synthetic data~\cite{Ring2018FlowbasedNT,engelmann2021conditional}, applicable for data augmentation in terms of class balancing.
In particular, some works claim that the tested generative models outperform classical oversampling techniques in terms of augmentation performance~\cite{Guo2021CombatingII,Liu2021AGA}.
In contrast, one work on systematic evaluation of classical resampling techniques in the NetFlow domain~\cite{Mogollon_2024} reports mixed results and emphasizes that none of the tested techniques reliably improves classification performance and can even decrease the performance of classifiers.  

In the domain of NIDS, existing studies either evaluate classical over- and undersampling methods without including modern generative models for data augmentation~\cite{Mogollon_2024}, or they conduct individual comparisons of generative models focusing solely on augmentation without incorporating classical sampling and cleaning techniques~\cite{Liu_Antypenko_Sushko_Zakharchenko_2022}. 
Moreover, the lack of standardization leads to individual generative models being tested on different datasets, each with distinct NetFlow features, which complicates direct comparisons of data generation approaches. 
Often, approaches are evaluated on a single dataset, raising questions about the generalizability of the results to other datasets. 
Furthermore, the arbitrarily chosen classifier models used to evaluate the resampling methods further complicates the overall comparability.

\begin{figure}[t]
    \centering
    \includegraphics[width=0.99\linewidth]{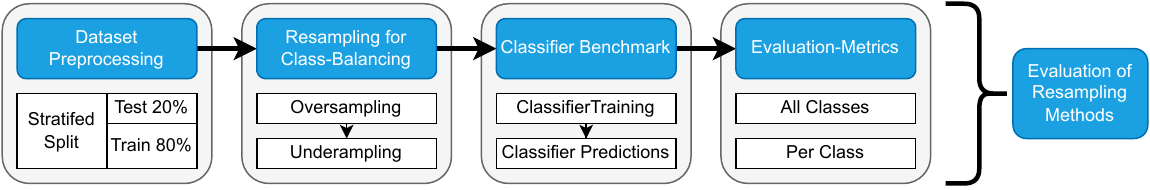}
    \caption{The main components of the experimental setup for the comparison of resampling methods}
    \label{fig:exp_setup}
\end{figure}

To address these gaps, our objective is to systematically evaluate the effectiveness of modern generative models to improve class balance through data augmentation. We compare these methods with classical (synthetic) oversampling techniques, both independently and in combination with undersampling methods. 
Specifically, we construct a testbed of various multi-class classifier models, which are evaluated using established classification metrics on NIDS benchmark datasets. This setup allows us to analyze the effects of both classical and modern resampling methods on different classification models. %TODO What are the reserach questions we want to answer?
An overview of our experimental setup is given in Figure \ref{fig:exp_setup}.

Our experiments generally reveal a minor negative impact of most resampling strategies on overall performance and rare classes in particular, while specific classifier-dependent combinations show improvements in classification performance.
This suggests that the arbitrary choice of resampling and augmentation techniques is not a reliable method for improving multi-class classification performance on NIDS data. Instead, this choice should be considered as hyperparameter and optimized separately for each specific classifier.

The key contributions in this paper are summarized as follows:
\begin{itemize}
    \item \emph{Comprehensive Evaluation Framework:} We present a systematic testbed incorporating multiple class-balancing strategies, including both classical resampling and modern generative approaches, in conjunction with various multi-class classification models.
    \item \emph{Comparative Analysis:} We compare the performance of 42 resampling combination on three datasets using various established classification metrics, ensuring a thorough analysis of the effects of different resampling methods.
    \item \emph{Reproducibility:} To promote reproducibility and facilitate further research, we have made our code\footnote{
    \url{https://github.com/maxwolf-code/netflow\_multiclass\_synthetic}} and experimental setup publicly available, allowing for a thorough quantitative evaluation of future NIDS data augmentation approaches.
\end{itemize}

\section{Related Work}
Various approaches for synthetic NetFlow generation and synthetic balancing can be found in literature.
Guo et al.~\cite{Guo2021CombatingII} use a Conditional GAN model to generate synthetic network data for oversampling on a single dataset called ISCXVPN2016 for multi-class classification. %cite
Their goal is to use synthetic data for oversampling minority classes, thereby enhancing the performance of classification models.
The setup of resampling methods consists of five traditional oversampling methods, namely Random Oversampling, SMOTE, ADASYN, SMOTE-SVM and SMOTE-TO\-MEK, which they compare to their CGAN approach.
Only one classifier, the 1D-CNN is used as the classification model in the evaluation, measuring the classification performance via F1-Score and AUC-PR.
The results show that their approach surpasses the other oversampling techniques, but their experiments are executed exclusively on a single dataset and model.

Liu et al.~\cite{Liu2021AGA} apply a WGAN-GP model for oversampling attack NetFlows. %not clear if multi-class or binary classficaton?
Their approach is tested against other oversampling methods such as SMOTE, ADASYN, K-SMOTE, G-SMOTE, ACGAN-SVM, and GAN in the datasets NSL-KDD~\cite{Tavallaee2009ADA}, UNSW-NB15~\cite{UNSW2015}, and CICIDS-2017~\cite{Sharafaldin2018TowardGA}. 
In their experiments they test seven classifiers which are Naive Bayes, Decision Tree, Random Forest, Gradient Boosting Decision Tree, Support Vector Machine, K-Nearest Neighbors and an Artificial Neural Network.
%They evaluate via standard the classification metrics accuracy, precision, recall and F1-Score. 
Their results suggest that the proposed oversampling method can effectively improve the detection performance.
However, all the tested approaches are based on interpolation-based approaches such as SMOTE variants and ADASYN or rely on GANs, allowing for only limited comparison to other state-of-the-art generative methods.
Moreover, their evaluation does not report baseline results of the classifiers when trained without resampled data. Therefore, the actual benefit of resampling remains unclear.

Liu et al.~\cite{Liu_Antypenko_Sushko_Zakharchenko_2022} apply a conditional Variational Autoencoder for oversampling.
In their experiments, they use the CIC-IDS-2018 dataset exclusively and incorporate two neural network models, a 1D-CNN and a GRU Network for attack classification. 
Their evaluation features a comparison of the classification performance of the models that are trained on the original data and the oversampled data. They report improved macro F1 scores by around 3\% and 5\%. 
While their work highlights the effect of class balancing in the NIDS domain, their results cannot be generalized since their findings rely on a single dataset and only two classification models.

Liu et al.~\cite{Jiang_Liu_2024} use a diffusion-based model called NetDiffusion for the generation of packet-based network data in the pcap format.
In their experiments, they use a self collected dataset consisting of video streaming, video conferencing and social media traffic.
They evaluate the generated synthetic data in various aspects, e.g., by statistical and machine learning performance evaluations, network analysis or syntax related checks by parsing via wireshark or tcp-replay.
In their findings, they report that the diffusion model generates high fidelity traffic with a high conformity to protocol specifications.
Moreover, they highlight that their diffusion model demonstrates potential for class balancing.
While their work focused on the generation of pcap traffic, their model is based on a diffusion model, which can be adapted for NetFlow data. We therefore adapt this model and include it in our comparison.

The most similar work to ours is~\cite{Mogollon_2024}, which examines the effects of oversampling combined with undersampling, albeit exclusively on the UNSW-NB15 dataset.
They report that there is no clear winning resampling combination that reliably improves classification performance. 
Although their work systematically tests the combination of resampling methods, they only evaluate on one dataset. Moreover, modern resampling techniques based on generative models are not tested alongside the classical approaches. 

In this study, our aim is to bridge this gap by conducting a comprehensive evaluation study that compares multiple combinations of over- and undersampling including generative models with various classifiers across three datasets.

\section{Experiments}
In this section, we begin by providing an overview of the resampling methods under evaluation, including classical methods for oversampling, generative models, and undersampling methods. Following this, we introduce the datasets and describe our experimental setup.

\subsection{Methods for Resampling}
An imbalanced data distribution refers to significant differences in the number of samples across classes, which can cause classifiers to exhibit poor discriminatory power, particularly concerning minority classes~\cite{Han_Kamber_Pei_2012}.

\noindent\textbf{Random Oversampling (ROS)} is a straightforward approach to addressing imbalanced class distributions by randomly sampling from the original distributions. This involves replicating samples from the minority class(es) and adding them to the dataset to increase the number of data points in those classes~\cite{Han_Kamber_Pei_2012}. 

\noindent\textbf{Synthetic Minority Over-sampling TEchnique (SMOTE),} introduced by Chawla et al.~\cite{chawla2002smote}, is an oversampling technique that addresses imbalanced class distributions by generating synthetic data points.
Unlike traditional methods that duplicate existing data, SMOTE creates new instances in the feature space using linear interpolations.
For each minority class, SMOTE identifies the $k$ nearest neighbors ($k$NN) from the same class.
Synthetic data points are then generated along the line segments connecting these neighbors.

\noindent\textbf{Adaptive synthetic oversampling (ADASYN)}~\cite{Haibo_2008} pursues to shift the decision boundary of a classifier towards samples that are difficult to learn, as they are close to datapoints of the majority class(es). 
The method is based on three main steps.
At first, the degree of class imbalance is calculated. 
Second, a $k$NN classifier is utilized to determine datapoints from the minority classes close to majority classes. 
Next, similar to SMOTE, synthetic data points between the minority and the majority are created.

\noindent\textbf{Conditional-Variational-Autoencoder (C-VAE)}~\cite{sohn2015learning} is a generative model, based on the Variational Autoencoder which consists of an encoder network and a Decoder network coupled via a latent space, typically represented by a parameterized Gaussian distribution.
The encoder $q(z\mid x)$ transforms data $x$ into the latent space via variables $z$.
The decoder $p(x\mid z)$ transforms the latent space $z$ back into the data space $x$.
The model is trained in a self-supervised manner for data reconstruction. 
The C-VAE extends this idea by conditioning the latent space on the output. 
Instead of assuming a standard Gaussian distribution for the prior of the latent representation, C-VAE uses the input observation to adjust the prior distribution of Gaussian latent variables $q(z\mid x,y)$ and the generation of the output by $p(y\mid x, z)$.
Our experiments use a C-VAE trained on multi-class labels of the NetFlow data, which allows to balance the imbalanced class distribution of the original data.

\noindent\textbf{Conditional-WGAN-GP (C-WGAN)}~\cite{Engelmann_Lessmann_2021} is based on Generative Adversarial Networks (GANs) which consist of two neural networks, a generator $G$ and a discriminator $D$.
Both neural networks play a min-max game, where $G$ tries to generate realistic fake data $X_f$ and $D$ tries to identify given data as real or fake.
Both models compete in this min-max objective, which trains the generator to generate realistic looking data from noise.

Conditional generation is performed by adding a class embedding $y$ as input to the generator and discriminator. Therefore, $G$ learns to model conditional data representations.
In our experiments, we use a C-WGAN-GP (Conditional Wasserstein Generative Adversarial Network with Gradient Penalty)~\cite{gulrajani2017improved,arjovsky2017wasserstein}, which is trained on multi-class NetFlow data to generate class-specific samples.

\noindent\textbf{Conditional Denoising Diffusion Probabilistic Model (C-DDPM)}~\cite{rombach2022high} incorporates a so-called diffusion process modeled as a Markov process. Diffusion-based models recently gained popularity in the image domain through high-quality image generation.
At first, images are compressed in a latent dimension via an encoder model that transforms the pixel space into a lower-dimensional continuous vector representation.
The latent representation can be decoded into the pixel space via a decoder.
Next, noise is added to the latent representation step by step, defined by time steps $t$.
Afterwards, a denoising U-Net neural network is trained to estimate the added noise for each time step $t$, which can be used to denoise the image. 
Eventually, the U-Net learns to denoise the latent representations, which allows to generate plausible data from randomly drawn noise input.
The conditioning, originally driven by a text prompt, directs the diffusion process to generate images according to the specified conditioning.
Our adapted diffusion model is trained on NetFlow data instead of images, and conditioned via class labels to generate class-specific data for oversampling.

%\subsection{Methods for Data Cleaning}

\noindent\textbf{Random Undersampling (RUS)}
samples data points from the minority class(es)  to reduce their quantity, aiming to balance the number of data points across classes~\cite{Han_Kamber_Pei_2012}.

\noindent\textbf{Tomek Links (Tomek)}~\cite{Tomek_1976}
are pairs of datapoints close to each other that belong to two different classes. 
In order to clean the decision boundaries, either both datapoints are deleted, or the datapoint of the majority classes is deleted. In our setup, we chose the removal of both data points.

\noindent\textbf{Edited Nearest Neighbor (ENN)}~\cite{Wilson_1972} is based on $k$NN datapoints that determine where the $k$-nearest datapoints of an observed datapoint belong to other classes. 
If so, there are two options for cleaning.
One option is to exclusively delete samples of the majority classes.
The second option is to delete all datapoints that are not similar to their $k$-nearest neighbors, which is applied in our experiments.

\noindent\textbf{Neighborhood Cleaning Rule (NCR)}~\cite{Laurikkala_2001}
combines two main aspects, the removal of redundant data points and the cleaning of the decision boundary. The redundant data points are identified via Condensed Nearest-Neighbor Clustering, where very similar datapoints are removed.
Next, based on a $k$NN-Classifier, ambiguous datapoints that are misclassified are removed to remove noisy data from the decision boundaries of the individual classes.

\noindent\textbf{Near Miss 3 (NN-3)}~\cite{Mani_Zhang_2003} is a heuristic cleaning procedure based on the Nearest Neighbor algorithm. 
NN-3 identifies the three nearest neighbors from the majority class for each example in the minority class and retains these nearest neighbors for training, thereby undersampling the majority class. 
By this heuristic, the algorithm keeps datapoints close to the borders of different classes that are most informative.

\subsection{Benchmark datasets}
The datasets applied in our experiments are originally based on packet-based datasets, which have been transformed via nProbe~\cite{nProbe} by Sarhan et al.~\cite{sarhan2021netflow}. 
While other NetFlow benchmark datasets lack a common NetFlow exporter and therefore a common format with identical features, the Flow features of these datasets are homogeneous, allowing to outrule the influence of different feature sets to potential outcomes.
%Especially machine learning based models, applied for NIDS, are sensitive towards input features and therefore a common set of features is mandatory for an accurate comparison.
%The NF-UNSW-NB15 mostly consists of normal traffic, while the NF-BoT-IoT and NF-ToN-IoT contain more malicious than normal traffic.
All datasets have a heavily imbalanced class distribution in common, especially in the multi-class setting.
%While some related work target the generation of attacks, these datasets incorporate the generation of attack as well as normal traffic in our multi-class setting.
%\\[1em]
                                                     
\noindent\textbf{NF-BoT-IoT}
is recorded botnet traffic data of a realistic environment. 
Ostinato and Node-red tools were utilized to generate the non-IoT and IoT traffic. %\cite{koroniotis2019towards}. 
The dataset contains 586\,241 (97.69\%) attack NetFlows and 13\,859 (2.31\%) benign NetFlows.
The attacks are dived into the listed types followed by the number of NetFlows:
Reconnaissance 470\,655, DDoS 56\,844, DoS 56\,833 and Theft 1\,909.

\noindent\textbf{NF-ToN-IoT}
features heterogeneous traffic based on telemetry data of Internet of Things (IoT) services, network traffic of IoT networks. % \cite{fesz-dm97-19}.
The dataset includes 1\,108\,995 (80.4\%) attack and 270\,279 (19.6\%) benign NetFlows.
The attack types and the corresponding number of samples are:
Backdoor 17\,247, DoS 17\,717, DDoS 326\,345, Injection 468\,539, MITM 1\,295, Password 156\,299, Ransomware 142, Scanning 21\,467 and XSS 99\,944

\noindent\textbf{NF-UNSW-NB15}
includes real network traffic and synthetic traffic based on XIA PerfectStorm. % \cite{moustafa2015unsw}.
This dataset has 1\,550\,712 (95.54\%) benign and 72\,406 (4.46\%) attack NetFlows.
The multi-class attack types are:
Fuzzers 19\,463, Analysis 1\,995, Backdoor 1\,782, DoS 5\,051, Exploits 24\,736, Generic 5\,570, Reconnaissance 12\,291, Shellcode 1\,365 and Worms 153.

\subsection{Experimental Setup}
Our setup comprises several multi-class classification models chosen for NetFlow classification based on current literature.
To be precise, we incorporate a broad range of popular and well-known models, specifically K-Nearest Neighbors ($k$NN)~\cite{Han_Kamber_Pei_2012}, Decision Tree (DTree)~\cite{Han_Kamber_Pei_2012}, Random Forest (RF)~\cite{Han_Kamber_Pei_2012}, Extra Tree (ExTree)~\cite{Geurts_Ernst_Wehenkel_2006}, Multi-Layer Perceptron (MLP)~\cite{Han_Kamber_Pei_2012}, and Extreme Gradient Boosting (XGBoost)~\cite{Chen_Guestrin_2016}.

We split each NIDS dataset into stratified train-test sets, with the training set comprising 80\% of the data and the test set containing the remaining 20\%. Given the extreme imbalance in the datasets, we generate stratified splits to ensure that each class is proportionally represented in both the training and test sets, maintaining the original class distributions.
In accordance with the publishers of the NIDS datasets~\cite{sarhan2021netflow}, we exclude IP addresses and ports, as they are strong indicators of attacks, given that certain attacks are often executed from similar IPs. 
Moreover, we exclude the L7\_PROTO field which contains Layer 7 protocol information which are sometimes unique for normal or attack behavior and are a label indicator as well.
Additionally, we normalize NetFlow attributes using min-max normalization based on the value ranges of each field of the NetFlow V9 specification \cite{nProbe}, before applying resampling methods and classification models.
The resampling methods are applied on the train set exclusively.

The objective comparison of classifier performance based on the resampling strategy requires quantitative evaluation metrics.
We report Weighted-Precision, Weighted-Recall, Weighted-F1-Score, and Mattheus Correlation Coefficient~\cite{Grandini_Bagli_Visani_2020} as they consider imbalanced class distributions to allow for an objective comparison.

\section{Results}
This section, describes and discusses the results of our study, first focusing on the overall classification performance, then on the individual class performance.

\subsection{Evaluation of the overall classification performance}

\begin{figure}[t]
    \centering
    \includegraphics[clip, trim=0cm 3.4cm 0cm 0cm,width=0.99\linewidth]{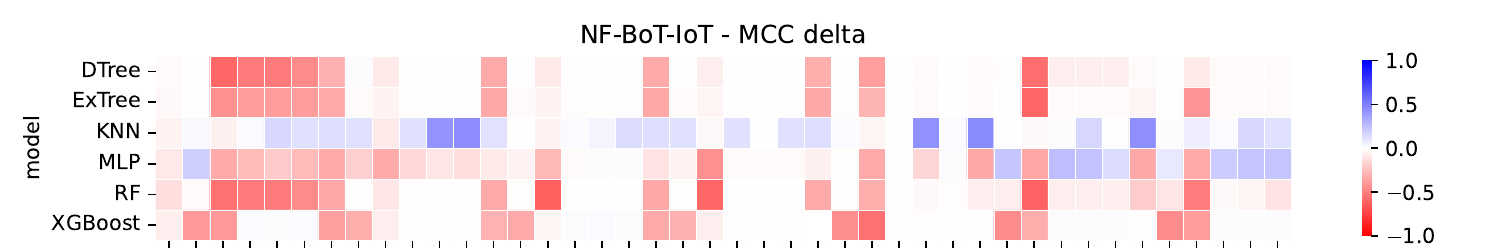}
    \includegraphics[clip, trim=0cm 3.4cm 0cm 0cm,width=0.99\linewidth]{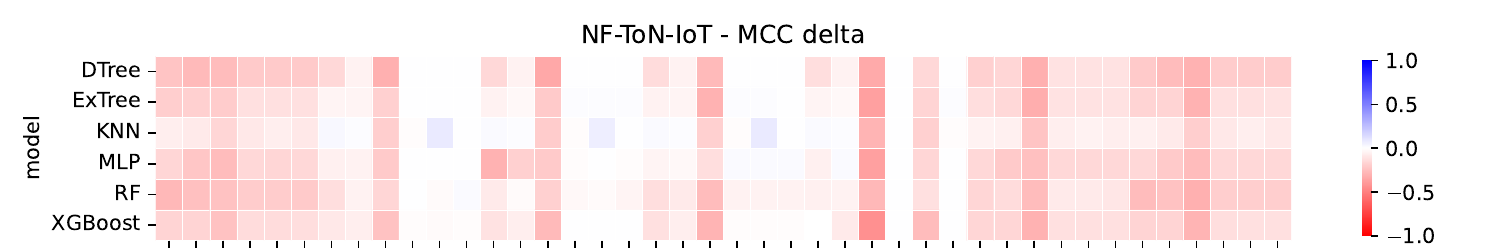}
    \includegraphics[width=0.99\linewidth]{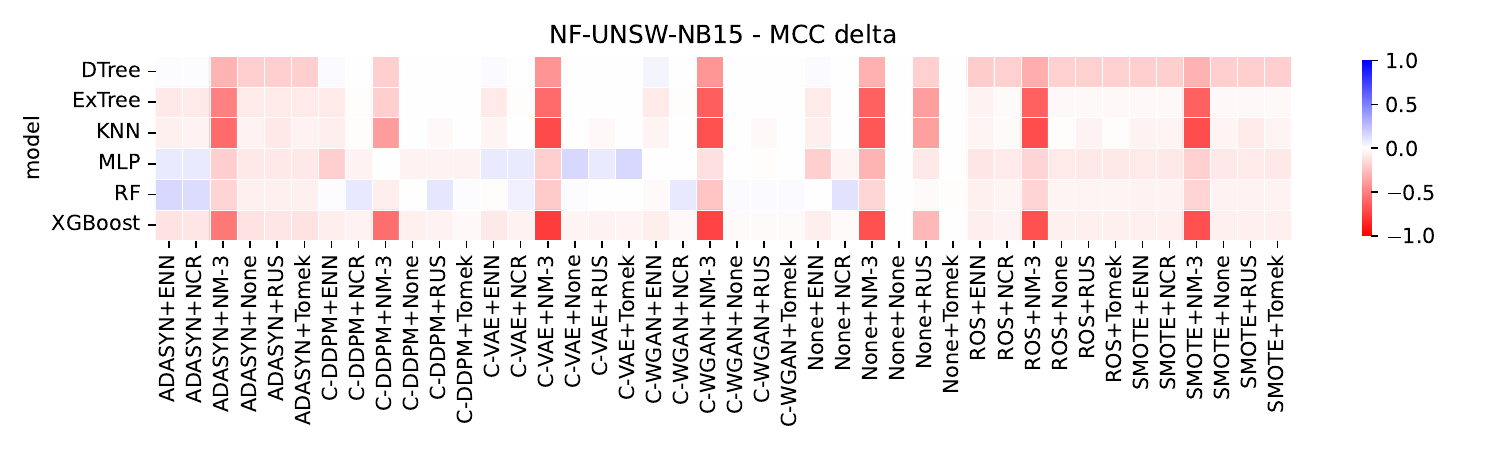}

    \caption{Delta MCC values per model: MCC of original (None) resampling - scores of resampling strategy on the y-axis.}
    \label{fig:delta_mcc}
\end{figure}

Our experiments reveal that some models show small improvements in classification performance when trained on resampled data. 
However, most of the results indicate lower performance compared to training on the original imbalanced training data across all tested classifier models.
The heatmaps in Figure~\ref{fig:delta_mcc} display the relative improvement in classification performance for each model trained on a specific resampling combination (oversampling + undersampling), calculated by subtracting the baseline MCC values (None + None) from each resampling combination score.
Blue color gradients highlight improved scores, red gradients indicate decreased scores, and white indicates no difference.

The heatmap demonstrates a minimal positive impact on the classification performance. 
There is no consistent improvement trend where a single resampling strategy consistently improves the results of a model across all datasets. 
For the NF-ToN-IoT dataset a notable improvement for the $k$NN and MLP model can be observed. However, this can be explained by the generally poor classification performance of these models trained on the imbalanced dataset. 
Consequently, resampling helps these models to achieve a classification performance comparable to other classifiers.
A similar pattern is seen with the MLP and RF classifiers on the NF-UNSW-NB15 dataset. Notably, some resampling methods, such as NM-3, consistently decrease performance across all datasets, suggesting that the undersampling process removes relevant information.

Next, we collected the resampling strategies with the best MCC scores for each model and dataset to analyze the total performance gain compared to the baseline reference where no resampling was applied, as shown in Table~\ref{tab:baseline_best}. The best values for each model and dataset are indicated in bold, while the best overall result for each dataset across all models is highlighted in blue. These results confirm the notable improvement mentioned above for the $k$NN and MLP models in the NF-Bot-IoT dataset, elevating their performance to match the MCCs of the other classification models.
In general, models that already achieve high scores can be observed to exhibit minimal improvement. Although one might infer that generative models dominate among the best results, it is important to highlight that our analysis focused exclusively on the top performing outcomes per MCC. This summary does not emphasize the general trend in which the models exhibited a decreased classification performance, as evident in Figure~\ref{fig:delta_mcc}.
Another major finding of Table~\ref{tab:baseline_best} is that the XGBoost model consistently yields the best or second best results among all datasets. This finding aligns with the literature, which indicates that XGBoost achieves comparable or superior performance to state-of-the-art models across diverse domains. Our findings suggest that testing different robust classifier models is more effective for achieving good classification results compared to employing different resampling techniques.

\begin{table}
\caption{Listing of the numeric codes per class}
    \label{tab:class_codes}
    \centering
    \resizebox{\textwidth}{!}{

    \begin{tabular}{l| lllll lllll}
    \toprule
    Num Code & 0 & 1 & 2 & 3 & 4 & 5 & 6 & 7 & 8 & 9 \\
    \midrule
Bot-IoT & Benign & Theft & DDoS & DoS & Reco.& --- &  --- & --- & --- & --- \\
ToN-IoT & Benign & DoS & Injection & DDos & Scanning & Password & MitM & XSS & Backdoor & Ransom. \\
UNSW-NB15  & Benign & Exploits & Reco. & DoS & Generic & Shellcode & Backdoor & Fuzzers & Worms & Analysis \\
    \bottomrule
    \end{tabular}
    }
\end{table}

\begin{table}
    \caption{Comparison of baseline (None+None) vs. the best  (MCC) results (in \%). Bold font indicates the best of all resampling results for the model and the given dataset. Blue color indicates the best results per dataset. Detection Rate (DR in \%) is given per Classes (c0 - c9), numeric class codes are explained in Table \ref{tab:class_codes}.}
    \label{tab:baseline_best}
%\tiny
\centering
\footnotesize

\resizebox{.99\textwidth}{!}{
    
\begin{tabular}{lll | rrrrr | rrr rrr rrrr}
\toprule
 
ds & model & sampling & Acc & Prec & Rec & F1 & MCC & \multicolumn{10}{c}{DR} \\

 & & & & & & & & c0 & c1 & c2 & c3 & c4 & c5 & c6 & c7 & c8 & c9 \\

\midrule
\multirow[t]{251}{*}{\rotatebox[origin=r]{90}{NF-BoT-IoT}} 
& \multirow[t]{42}{*}{DTree} 
   & None+None & 86.304 & 84.413 & 86.304 & 84.813 & 59.596 & 80 & 3 & 48 & 24 & 98 & --- & --- & --- & --- & --- \\
 &  & C-DDPM+NCR & 86.366 & 83.425 & 86.366 & 83.841 & \textbf{60.445} & 87 & 2 & 7 & 64 & 98 & --- & --- & --- & --- & --- \\
\cline{2-18}
 & \multirow[t]{42}{*}{ExTree}
    & None+None & 83.796 & 88.433 & 83.796 & 83.212 & 62.407 & 82 & 3 & 93 & 5 & 92 & --- & --- & --- & --- & --- \\
 &  & ADASYN+NCR & 83.629 & 88.030 & 83.629 & 83.312 & \textcolor{blue}{\textbf{62.516}} & 91 & 5 & 93 & 6 & 91 & --- & --- & --- & --- & --- \\
\cline{2-18}
 & \multirow[t]{42}{*}{$k$NN}
   & None+None & 28.654 & 82.492 & 28.654 & 33.936 & 15.471 & 90 & 4 & 48 & 51 & 21 & --- & --- & --- & --- & --- \\
 &  & ROS+None & 83.456 & 85.731 & 83.456 & 81.943 & \textbf{61.009} & 91 & 4 & 91 & 8 & 21 & --- & --- & --- & --- & --- \\
\cline{2-18}
 & \multirow[t]{42}{*}{MLP}
   & None+None & 80.870 & 70.666 & 80.870 & 74.916 & 33.197 & 70 & 0 & 18 & 0 & 98 & --- & --- & --- & --- & --- \\
 &  & ROS+None & 80.829 & 87.007 & 80.829 & \textbf{81.299} & \textbf{58.403} & 84 & 6 & 5 & 94 & 88 & --- & --- & --- & --- & --- \\
\cline{2-18}
 & \multirow[t]{42}{*}{RF} 
    & None+None & 86.342 & 84.216 & 86.342 & 83.792 & 60.083 & 79 & 2 & 8 & 63 & 98 & --- & --- & --- & --- & --- \\
 &  & C-VAE+RUS & 86.506 & 84.196 & 86.506 & 83.802 & \textbf{60.453} & 79 & 2 & 64 & 7 & 99 & --- & --- & --- & --- & --- \\
\cline{2-18}
 & \multirow[t]{41}{*}{XGBoost}
    & None+None & 83.828 & 87.248 & 83.828 & 85.033 & 60.705 & 84 & 4 & 27 & 72 & 92 & --- & --- & --- & --- & --- \\
 &  & C-VAE+RUS & 83.817 & 88.138 & 83.817 & 83.282 & \textbf{62.469} & 83 & 3 & 93 & 6 & 92 & --- & --- & --- & --- & --- \\
%\cline{1-13} \cline{2-18}

%\cline{1-13} \cline{2-18}
\midrule
\multirow[t]{251}{*}{\rotatebox[origin=r]{90}{NF-ToN-IoT}} & \multirow[t]{42}{*}{DTree}

   & None+None & 49.891 & 43.228 & 49.891 & \textbf{41.548} & 33.991 & 68 & 0 & 89 & 19 & 0 & 0 & 12 & 0 & 94 & 3 \\
 &  & C-DDPM+Tomek & \textbf{49.9} & 43.239 & \textbf{49.9} & 41.547 & \textbf{34.013} & 68 & 0 & 89 & 19 & 0 & 0 & 12 & 0 & 94 & 3 \\
\cline{2-18}
 & \multirow[t]{42}{*}{ExTree} 
   & None+None & 52.476 & 45.755 & 52.476 & 44.020 & 38.242 & 72 & 0 & 95 & 19 & 0 & 0 & 13 & 0 & 94 & 3 \\
  &  & None+Tomek & 52.760 & 46.641 & 52.760 & \textbf{44.24} & \textbf{38.834} & 74 & 0 & 95 & 19 & 0 & 0 & 13 & 0 & 94 & 3 \\
\cline{2-18}
 & \multirow[t]{42}{*}{$k$NN} 
   & None+None & 48.053 & 50.256 & 48.053 & 41.219 & 32.908 & 86 & 1 & 82 & 3 & 3 & 8 & 35 & 0 & 94 & 96 \\
 &  & C-DDPM+RUS & \textbf{52.251} & 50.509 & \textbf{52.251} & 41.583 & \textbf{40.693} & 87 & 46 & 95 & 3 & 0 & 1 & 35 & 0 & 94 & 96 \\
\cline{2-18}
 & \multirow[t]{42}{*}{MLP}
   & None+None & 45.682 & 37.772 & 45.682 & 38.051 & 26.411 & 60 & 0 & 81 & 20 & 0 & 0 & 0 & 0 & 94 & 0 \\
   &  & C-WGAN+None & \textbf{46.613} & 39.920 & \textbf{46.613} & \textbf{39.18} & \textbf{28.188} & 55 & 0 & 86 & 21 & 0 & 0 & 0 & 0 & 94 & 0 \\
\cline{2-18}
 & \multirow[t]{42}{*}{RF}
   & None+None & 48.761 & 40.640 & 48.761 & 40.763 & 31.492 & 68 & 0 & 85 & 21 & 0 & 0 & 6 & 0 & 94 & 3 \\
 & & C-DDPM+Tomek & \textbf{49.829} & 43.184 & \textbf{49.829} & \textbf{41.497} & \textbf{33.88} & 68 & 0 & 89 & 19 & 0 & 0 & 6 & 0 & 94 & 3 \\
 \cline{2-18}
 & \multirow[t]{41}{*}{XGBoost}
   & None+None & \textbf{56.51} & 54.627 & \textbf{56.51} & 47.883 & 45.987 & 92 & 0 & 96 & 18 & 0 & 1 & 44 & 0 & 94 & 96 \\
  &  & C-VAE+None & 56.339 & 53.467 & 56.339 & 47.416 & \textcolor{blue}{\textbf{46.011}} & 89 & 0 & 97 & 18 & 0 & 0 & 18 & 0 & 94 & 96 \\

%\cline{1-13} \cline{2-18}
\midrule
\multirow[t]{252}{*}{\rotatebox[origin=r]{90}{NF-UNSW-NB15}} & \multirow[t]{42}{*}{DTree}
& None+None & 95.812 & 93.753 & 95.812 & 94.356 & 30.901 & 99 & 19 & 0 & 4 & 12 & 0 & 0 & 3 & 12 & 0 \\
 &  & C-WGAN+ENN & 94.509 & 95.338 & 94.509 & 94.312 & \textbf{34.471} & 98 & 15 & 50 & 2 & 8 & 0 & 16 & 4 & 6 & 0 \\
\cline{2-18}
 & \multirow[t]{42}{*}{ExTree}
  & None+None & 97.009 & 96.612 & 97.009 & 96.502 & 62.027 & 99 & 85 & 19 & 16 & 40 & 0 & 2 & 26 & 12 & 0 \\
 &  & C-DDPM+Tomek & \textbf{97.013} & 96.595 & \textbf{97.013} & \textbf{96.564} & \textbf{62.508} & 99 & 85 & 28 & 15 & 38 & 0 & 3 & 26 & 12 & 0 \\
\cline{2-18}
 & \multirow[t]{42}{*}{$k$NN}
 & None+None & 96.992 & 97.289 & 96.992 & 97.074 & 67.749 & 98 & 78 & 78 & 26 & 46 & 43 & 8 & 50 & 38 & 6 \\
  &  & C-VAE+Tomek & 97.011 & 97.272 & 97.011 & \textbf{97.101} & \textbf{67.896} & 98 & 79 & 78 & 26 & 46 & 44 & 10 & 50 & 45 & 7 \\
 &  & C-WGAN+Tomek & 97.006 & 97.260 & 97.006 & 97.084 & \textbf{67.896} & 98 & 79 & 78 & 25 & 46 & 43 & 10 & 50 & 45 & 3 \\
\cline{2-18}
 & \multirow[t]{42}{*}{MLP}
 & None+None& \textbf{95.601} & 92.081 & \textbf{95.601} & \textbf{93.732} & 19.013  & 99 & 3 & 1 & 0 & 0 & 0 & 0 & 0 & 0 & 6 \\
  &  & C-VAE+Tomek & 93.603 & 95.214 & 93.603 & 93.620 & \textbf{33.998} & 97 & 0 & 0 & 0 & 0 & 0 & 10 & 62 & 0 & 64 \\
\cline{2-18}
 & \multirow[t]{42}{*}{RF}
 & None+None & 95.638 & 93.938 & 95.638 & 93.751 & 16.626 & 99 & 8 & 0 & 1 & 2 & 0 & 0 & 3 & 9 & 0 \\
 & & ADASYN+ENN& 93.829 & 95.791 & 93.829 & 93.989 & \textbf{31.552} & 97 & 5 & 0 & 2 & 0 & 0 & 47 & 0 & 83 & 39 \\
 \cline{2-18}
 & \multirow[t]{42}{*}{XGBoost}
  & None+None & 97.406 & 97.225 & 97.406 & \textbf{97.213} & \textcolor{blue}{\textbf{69.53}} & 99 & 84 & 75 & 18 & 46 & 13 & 10 & 32 & 12 & 0 \\
\bottomrule
\end{tabular}
}
\end{table}

\subsection{Evaluation of the classification performance per class}
\begin{figure}[h!]
    \centering
    \includegraphics[clip, trim=0cm 0.8cm 0cm 0cm,width=0.32\linewidth]{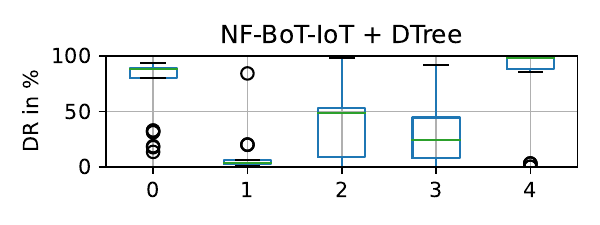}
    \includegraphics[clip, trim=0cm 0.8cm 0cm 0cm,width=0.32\linewidth]{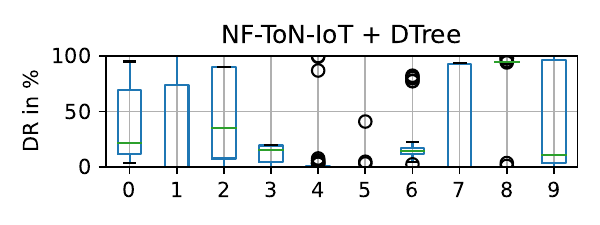}
    \includegraphics[clip, trim=0cm 0.8cm 0cm 0cm,width=0.32\linewidth]{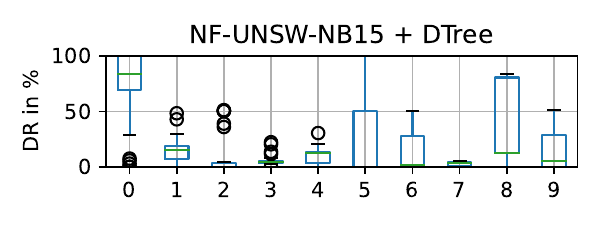}
    
    \includegraphics[clip, trim=0cm 0.8cm 0cm 0cm,width=0.32\linewidth]{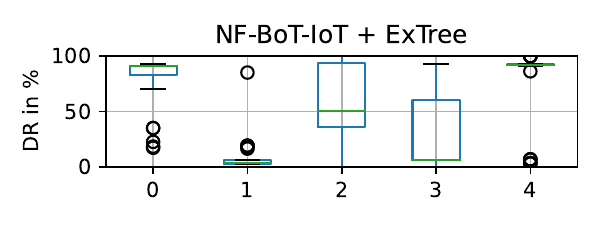}
    \includegraphics[clip, trim=0cm 0.8cm 0cm 0cm,width=0.32\linewidth]{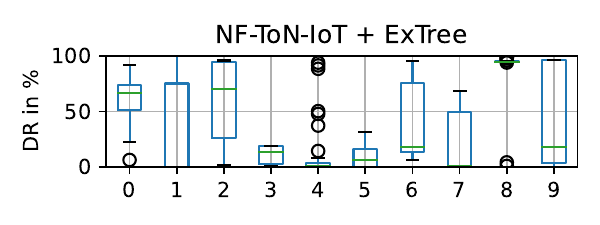}
    \includegraphics[clip, trim=0cm 0.8cm 0cm 0cm,width=0.32\linewidth]{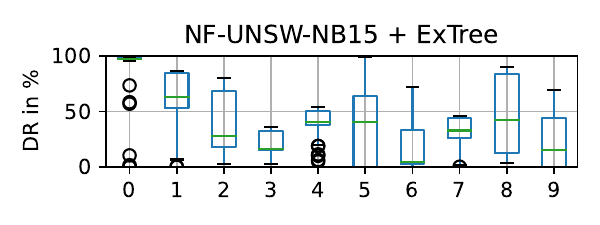}

    \includegraphics[clip, trim=0cm 0.8cm 0cm 0cm,width=0.32\linewidth]{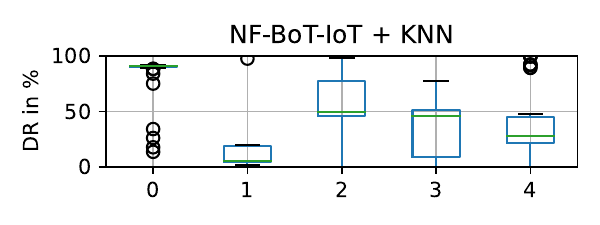}
    \includegraphics[clip, trim=0cm 0.8cm 0cm 0cm,width=0.32\linewidth]{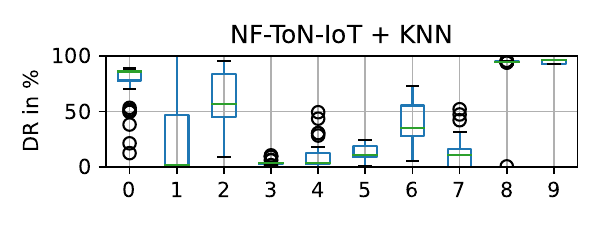}
    \includegraphics[clip, trim=0cm 0.8cm 0cm 0cm,width=0.32\linewidth]{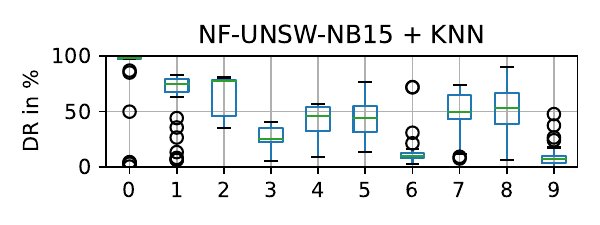}

    \includegraphics[clip, trim=0cm 0.8cm 0cm 0cm,width=0.32\linewidth]{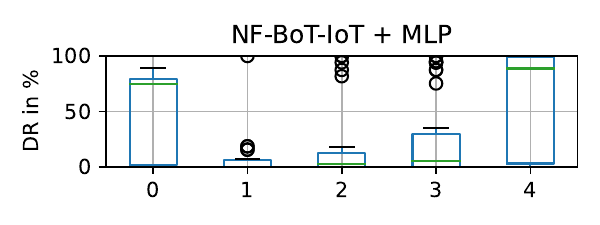}
    \includegraphics[clip, trim=0cm 0.8cm 0cm 0cm,width=0.32\linewidth]{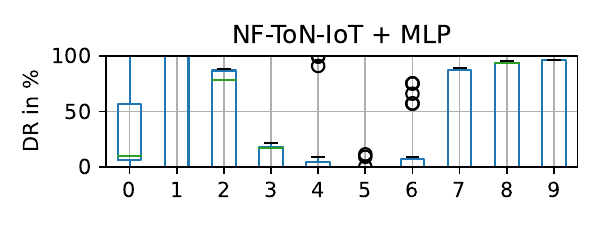}
    \includegraphics[clip, trim=0cm 0.8cm 0cm 0cm,width=0.32\linewidth]{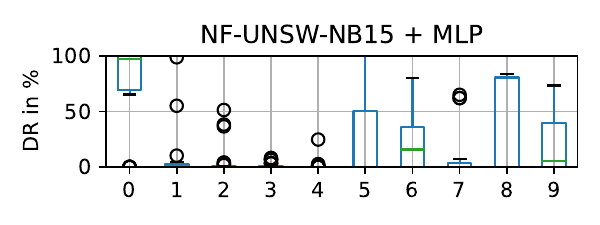}

    \includegraphics[clip, trim=0cm 0.8cm 0cm 0cm,width=0.32\linewidth]{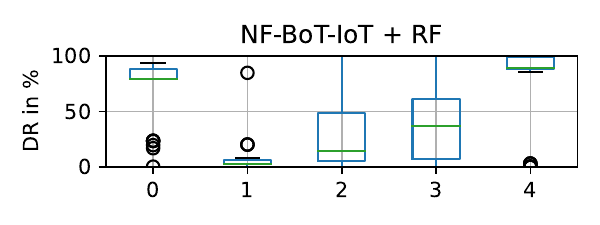}
    \includegraphics[clip, trim=0cm 0.8cm 0cm 0cm,width=0.32\linewidth]{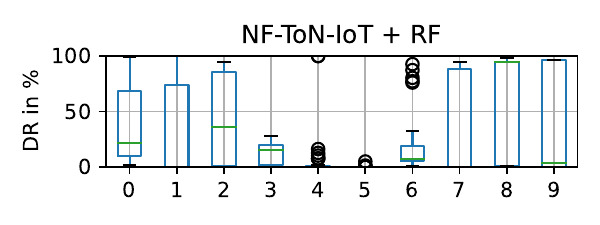}
    \includegraphics[clip, trim=0cm 0.8cm 0cm 0cm,width=0.32\linewidth]{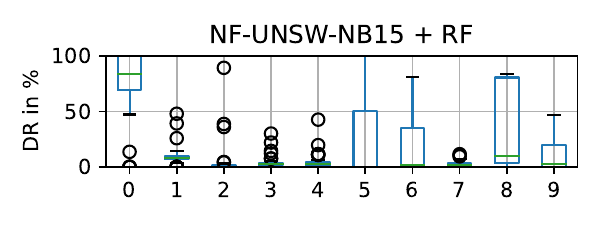}

    \includegraphics[width=0.32\linewidth]{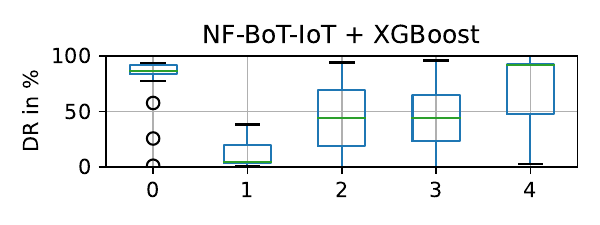}
    \includegraphics[width=0.32\linewidth]{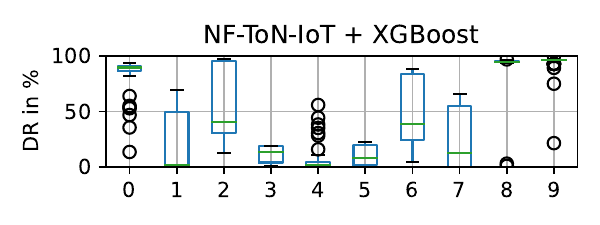}
    \includegraphics[width=0.32\linewidth]{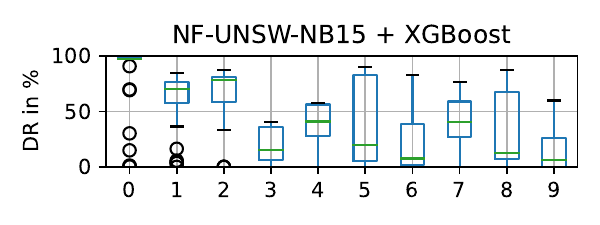}
    \caption{Metric scores plotted as mean values of all models. (numeric class codes explained in Table \ref{tab:class_codes}}
    \label{fig:boxplot_class}
\end{figure}

In this section, we want to further analyze the improvement of the multi-class classification performance for each model.
First, the classification performance per class using the detection rate is analyzed. 
The boxplots in Figure~\ref{fig:boxplot_class} show the distribution of detection rates (y-axis) per class (x-axis) among all models and applied resampling combinations.
One can observe classes which can be classified reliably by many models e.g., Class 0 of NF-BoT-IoT. In contrast, classes like Class 3 and 5 of NF-ToN-IoT could not be classified correctly, and resampling did not improve the performance in these cases.
Between these extremes lie classes such as Class 1, 2, and 9 of NF-ToN-IoT, which exhibit a wide range of detection rates.
This suggests a strong influence of resampling for some classifiers,
since e.g. Fig. \ref{fig:boxplot_class} shows a wide range of results for the class 9 of NF-ToN-IoT for some classifier models like DTree but not for others like KNN or XGBoost.

After analyzing the overall impact on per-class performance, we proceeded to examine specific improvements across different models and datasets. Table~\ref{tab:baseline_best} presents the detection rates per class for both the baseline and the best results. Similarly to the findings depicted in the boxplots in Figure~\ref{fig:boxplot_class}, certain classes cannot be correctly classified by both the baseline models and the best models optimized by resampling.
Although there are instances of improvements, for example $k$NN for NF-ToN-IoT in c1, where the baseline detection rate is strongly improved, these cases are exceptional.
Overall, our analysis suggests that resampling may not reliably enhance classification performance for a given NIDS dataset, although some models benefit from the right choice of resampling techniques.
A notable mention is XGBoost applied to the NF-UNSW-NB15 dataset, which performed best with the baseline data distribution across a broad spectrum of classes, although it did not achieve peak performance for each individual detection rate per class.   
While Fig.~\ref{fig:boxplot_class} demonstrates that some classes are challenging to classify in most cases, there are notable outliers with high performance, such as class 1 in NF-BoT-IoT or class 4 in NF-ToN-IoT.
Certain resampling methods, such as NM-3, can enhance the performance for smaller, harder-to-classify classes. However, this improvement often comes at the cost of significantly reducing the detection rates for other classes, thereby lowering the overall performance, as also illustrated in Fig.~\ref{fig:delta_mcc}.

\section{Limitations}
In this section, we acknowledge the limitations of our systematic evaluation of resampling methods.
Due to the resource-intensive nature of testing multiple resampled training datesets, we selected a set of efficient and established classification models omitting some newer approaches such as E-Graph-Sage~\cite{lo2022graphsage} in favor of XGBoost, which demonstrates comparable classification performance. Also, extensive hyperparameter evaluations per sampling combination could not be conducted due to computational limitations that might be eventually influenced by the change in class distributions. However, we believe that a fixed hyperparameter set per model allows for a fair comparison of the sampling strategies without, for example, changing the model capacity.
There exist other forms of sampling bias, namely temporal bias \cite{Pendlebury2018TESSERACTEE} which were out of scope of our experimental setup, since the tested datasets do not contain timestamps.
Additionally, in the context of generative models, we had to adapt them to our datasets and therefore could not test the exact same models as they were partially trained on different data such as pcap files. 
Lastly, we highlight the need for thorough quality checks on generated data. Our study focused solely on assessing the impact of generated data for data augmentation. However, there are broader concerns regarding the quality of synthetic data, such as the presence of unrealistic data or the potential leakage of original data in synthetic datasets created by models that replicate training data too closely.

\section{Conclusion}
This work aimed to systematically analyze the influence of resampling methods for class balancing on multi-class classification models.
The review of related works revealed a lack of standardization, which hinders the comparison of resampling approaches.
We compiled a setup of multiple classical and state-of-the-art resampling strategies in conjunction with established multi-class classification models.
Overall, we tested six classification models on 42 resampling combinations using three datasets.
Our assessment of classification performance considers both overall performance and performance per class.
Our results revealed a low positive impact of resampling on overall classification performance.
The majority of resampling strategies had a negative impact on the classification performance across all models, suggesting that arbitrary resampling may not reliably improve multi-class classification performance.
The per-class performance analysis indicated that the classification performance per class generally did not improve, as certain classes could not be classified well even after resampling.
Even the resampling combinations that exhibited the greatest improvements in overall classification performance did not enhance the detection rates of certain rare classes.
In conclusion, our findings suggest prioritizing the testing of different classification models, given the unreliable nature of improvements in resampling performance.
In the future, our aim is to expand our study to include additional datasets and classification models. 
Moreover, conducting a data-driven analysis of the augmented datasets, e.g., via a NetFlow evaluation framework~\cite{WOLF2024103993} is necessary to identify the factors contributing to both improved and diminished classification performances.

% ---- Bibliography ----
%
% BibTeX users should specify bibliography style 'splncs04'.
% References will then be sorted and formatted in the correct style.
%
\begin{credits}
\subsubsection{Acknowledgements} 
This work is funded by the Bavarian Ministry of Economic Affairs, Regional Development and Energy through the GENESIS project (grant no. DIK0422/03).
\end{credits}

\bibliographystyle{splncs04}
\bibliography{bib}

\end{document}